\documentclass[twocolumn]{elsart3}
\usepackage{graphics}
\usepackage{graphicx}
\usepackage{epsfig}
\usepackage{subfigure}
\usepackage{amssymb}

\begin{document}

\begin{frontmatter}



\title{First experiences with the\\ATLAS Pixel Detector Control System\\ at the Combined Test Beam 2004}


%\author[Wuppertal]{Martin Imh\"{a}user},
%\ead{imhaeuse@physik.uni-wuppertal.de}
%\author[Wuppertal]{Karl-Heinz Becks},
%\author[Wuppertal]{Tobias Hen\ss},
%\author[Wuppertal]{Susanne Kersten},
%\author[Wuppertal]{Peter M\"{a}ttig} and
%\author[Wuppertal]{Joachim Schultes}
%\address[Wuppertal]{University of Wuppertal, Gau\ss stra\ss e 20, 42119 Wuppertal, Germany}

\author{Martin Imh\"{a}user},
\ead{imhaeuse@physik.uni-wuppertal.de}
\author{Karl-Heinz Becks},
\author{Tobias Hen\ss},
\author{Susanne Kersten},
\author{Peter M\"{a}ttig} and
\author{Joachim Schultes}

\address{University of Wuppertal, Gau\ss stra\ss e 20, 42119 Wuppertal, Germany}

\begin{abstract}
% Text of abstract
\\Detector control systems (DCS) include the read out, control and supervision of hardware devices as well as the monitoring of external systems like cooling system and the processing of control data. The implementation of such a system in the final experiment has also to provide the communication with the trigger and data acquisition system (TDAQ). In addition, conditions data which describe the status of the pixel detector modules and their environment must be logged and stored in a common LHC wide database system.\\
At the combined test beam all ATLAS subdetectors were operated together for the first time over a longer period. To ensure the functionality of the pixel detector a control system was set up.\\  
We describe the architecture chosen for the pixel detector control system, the interfaces to hardware devices, the interfaces to the users and the performance of our system. The embedding of the DCS in the common infrastructure of the combined test beam and also its communication with surrounding systems will be discussed in some detail.
\end{abstract}

\end{frontmatter}

% main text
%%%%%%%%%%
\section{Introduction}
\label{sec:Intro}
%%%%%
For the first time, segments of all ATLAS subdetectors were integrated and operated together with a common Trigger and Data Acquisition (TDAQ), close to final electronics and the Detector Control System (DCS) at a CERN test beam.\\
During this test and certainly in the future experiment the overall aim of the DCS was and is to guarantee a reliable physics data taking and a safe operation of the detector. This is done by monitoring and controlling the DCS hardware, reacting to error conditions, providing several user interfaces and maintaining the communication to common infrastructure of the ATLAS experiment like TDAQ or database systems. Especially the communication between TDAQ and DCS is of major importance for the operation of the pixel detector as tuning of the read out chain requires access to both systems in parallel.
%%%%%%%%%%%
\section{Experimental Setup}
\label{sec:setup}
%%%%%
\begin{figure}[htb]
  \centering
      \subfigure{
          \includegraphics[width=\columnwidth]{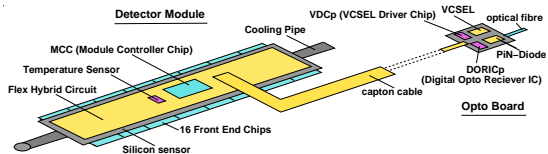}
          			}
  \caption{Pixel detector module}
  \label{fig:modul}
\end{figure}
\begin{figure}[h]
	\centering
		\includegraphics[width=0.7\columnwidth]{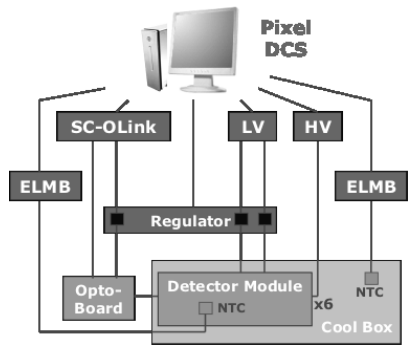}
	\caption{Pixel DCS test beam set up}
	\label{fig:Setup}
\end{figure}

 
The pixel detector module, shown in figure \ref{fig:modul}, is the smallest unit pixel DCS can act on. It consists of a silicon sensor and 16 front end chips as well as a Module Controller Chip (MCC) gathering hit data and servicing trigger requests.\\
Every detector module is connected to an optoboard which converts the electrical data signals transmitted from the modules to an optical signal for transmission to the off-detector electronics via optical fibres. In parallel it receives optical signals from the off-detector electronics and converts these to electrical signals for distribution to the modules.\\
The off-detector component Back Of Crate card (BOC) which serves as the optical interface between the Read Out Driver (ROD) and the optoboard \cite{Flick} is controlled by TDAQ while DCS takes care of the on-detector component optoboard.\\  
To operate six pixel detector modules as a part of the whole pixel detector its DCS provided various equipment at the combined test beam (shown in figure \ref{fig:Setup}). For more details about the design of the pixel DCS please refer to \cite{Imhaeuser}.\\
General purpose IO devices for the read out of the DCS hardware (ELMB\footnote{\textbf{E}mbedded \textbf{L}ocal \textbf{M}onitor \textbf{B}oard}) developed by  ATLAS DCS group, a home made supply system of three low voltage sources together with a reset signal to operate the optoboard (SC-OLink\footnote{\textbf{S}upply and \textbf{C}ontrol for the \textbf{O}pto\textbf{Link}}), a regulator system for protecting the FE chips of the detector modules, developed by INFN Milano, a high voltage source for the depletion of the sensors and also temperature and humidity sensors have come into operation. To integrate the hardware and to supervise the system in-house developed software tools have been used.
%%%%%%%%%%%%
\section{Detector Control System}
\label{sec:DCS}
The ATLAS detector is hierarchically organized in a tree-like structure into sudetectors, sub-systems, etc.. This has to be reflected in the design and implementation of the DCS.\par
\begin{figure}[h]
	\centering
		\includegraphics[width=\columnwidth]{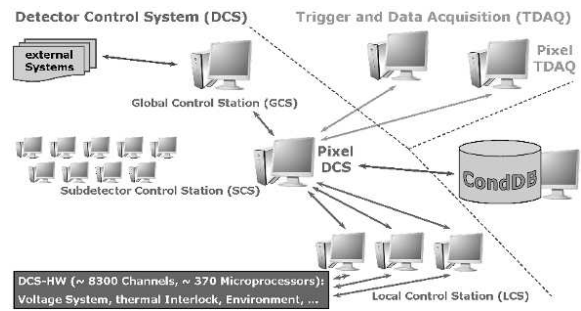}
	\caption{Detector control system}
	\label{fig:DCS}
\end{figure}
Therefore DCS is organized in three functional layers:
\begin{itemize}
\item{the \textit{global control station} which e.g. provides tools for the overall operation of ATLAS,}
\item{the \textit{subdetector control station} which e.g. provides full stand-alone control capability and synchronises the supervision of all subsystems below and}
\item{the \textit{local control station} which e.g. reads data from the DCS hardware.}
\end{itemize}
The core of the software is based on the commercial Supervisory Control And Data Acquisition (SCADA) package PVSS\footnote{\textbf{P}roze\ss - \textbf{V}isualisierungs und \textbf{S}teuerungs- \textbf{S}oftware, ETM, Austria}. PVSS allows to gather information from the DCS hardware and offers the implementation of supervisory control functions such as data processing, alert handling and trending. It has a modular architecture based on functional units called managers. Applications can be distributed over many stations on the network which defines a distributed system \cite{Burckhart}.\\
%%%%%%%
\subsection{Distributed System}
\label{subsec:distri}
%%%
\begin{figure}[htb]
	\centering
		\includegraphics[width=0.5\columnwidth]{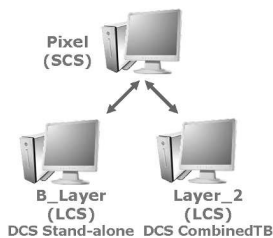}
	\caption{Distributed system}
	\label{fig:Distributed}
\end{figure}
%%%
%\begin{figure*}[htb]
%  \centering
%      \subfigure{
%          \includegraphics[width=2\columnwidth]{corel_bilder/Bild31.eps}
%          			}
%  \caption{Graphical user interface - SIT}
%  \label{fig:SIT}
%\end{figure*} 
At the combined test beam we embedded three PVSS stations as a distributed system based on a full detector simulation as shown in figure \ref{fig:Distributed}. This test demonstrated successfully the partitioning over several computers and their interconnection in a common environment.\par
%%%%%%%
\subsection{Software tools}
\label{subsec:tools}
\begin{figure*}[h]
	\centering
		\includegraphics[width=0.9\textwidth]{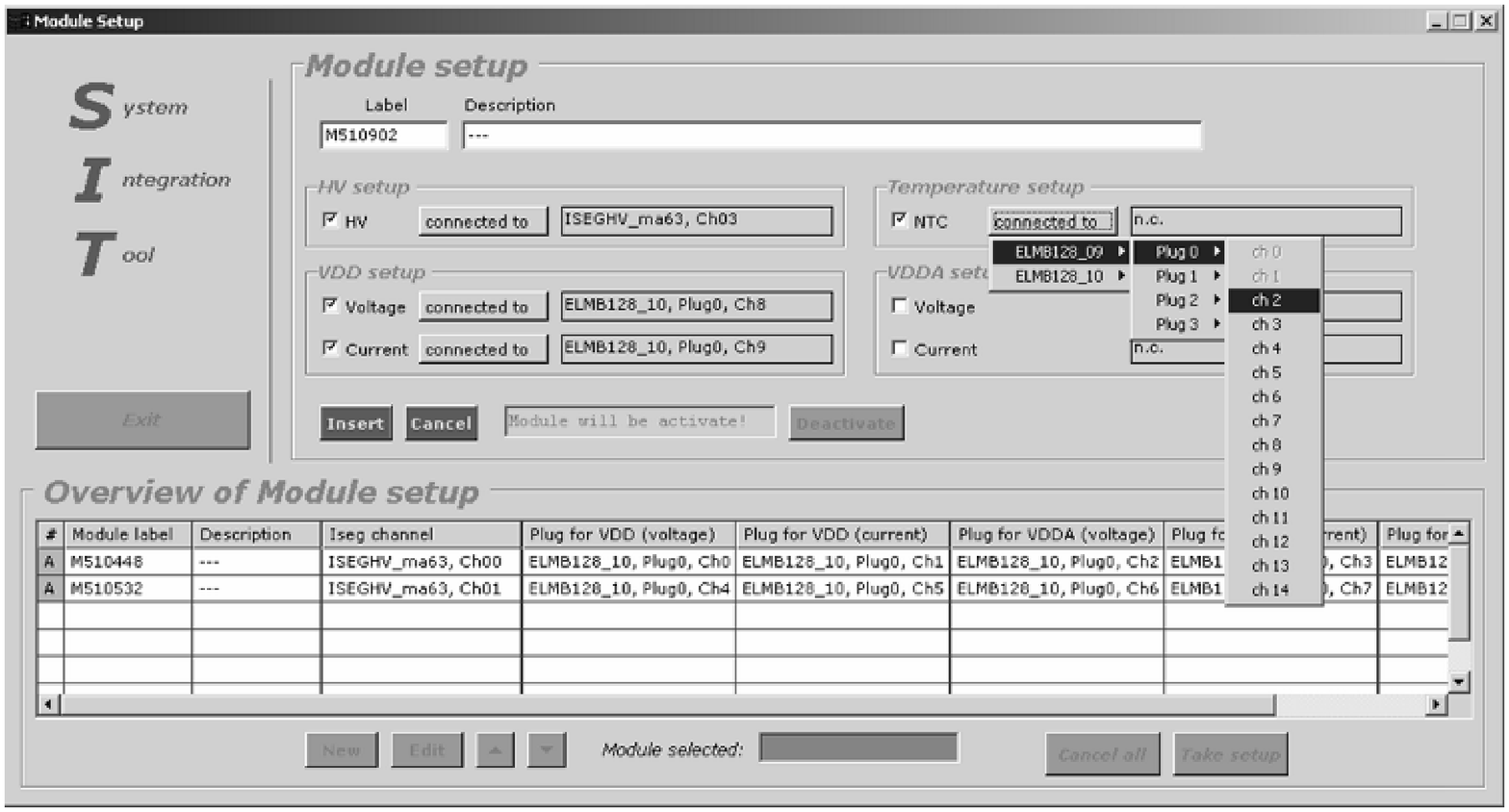}
	\caption{Graphical user interface - SIT}
	\label{fig:SIT}
\end{figure*}

%%%%%
The software of the pixel DCS consists of several subprojects such as tools for the implementation of the DCS hardware in the software environment and the configuration in an automated way, tools for combining all information concerning one detector module in a flexible way (see figure \ref{fig:SIT}, last page) and also graphical user interfaces. For example figure \ref{fig:SIT} (last page) shows the System Integration Tool (SIT) which follows the detector hierarchy and therefore maps the real cabling structure into the software.\\
%%
%\begin{figure*}[b]
%	\centering
%%		\includegraphics[width=\columnwidth]{corel_bilder/sit.eps}
%		\includegraphics[width=2\columnwidth]{corel_bilder/Bild31.eps}
%	\caption{Graphical user interface - SIT}
%	\label{fig:SIT}
%\end{figure*}
All these software tools were used at the combined test beam and the experience now helps to develop advanced tools for the experiment.\\
%%%%%%%%%%
\section{DAQ-DCS Communication}
\label{section:DDC}
\begin{figure}[h]
	\centering
				\includegraphics[width=\columnwidth]{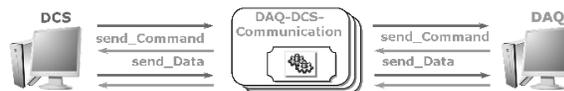}
	\caption{Schematic of DDC}
	\label{fig:DDC}
\end{figure}

%%%%%
TDAQ and DCS are controlled by finite state machines which consist of different states and transition functions which map a start state to a next state. Both systems are independent while TDAQ has the master control during detector operation. This means that the TDAQ finite state machine has to be able to cause transitions in the DCS finite state machine. Further more TDAQ applications have to transfer selective data to DCS as well as DCS must make required data available to  TDAQ. Nevertheless TDAQ must be informed about state conditions.\\
To cover all the required transfers, the DAQ-DCS Communication (DDC) software \cite{Khomoutnikov} has been developed by the ATLAS DCS group (see figure \ref{fig:DDC}).\par 
%%%
%%%
%\begin{figure*}[h]
%	\centering
%%		\includegraphics[width=\columnwidth]{corel_bilder/DDCtest.eps}
%		\includegraphics[width=2\columnwidth]{corel_bilder/Bild4.eps}
%	\caption{Graphical user interface - DDC}
%	\label{fig:ddctest}
%\end{figure*}
DDC was set up in pixel configuration by the authors and was running for four months in the combined environment. During this time the pixel specific DDC application was tested intensely.\\
%%%
Concerning the command transfer, we were able to show that the used pixel DCS finite state machine reacted in a well defined way on TDAQ transitions. Additionally pixel DCS directly computed actions via DDC in response to three TDAQ transitions at the combined test beam. Further more the possibility to set DCS hardware with TDAQ applications without changing the TDAQ state was tested successfully.\par
Regarding the data transfer, DCS visualised data like temperatures, depletion and low voltages of the detector modules or the states of DCS hardware for TDAQ while DCS received data from TDAQ like the status of TDAQ services or run parameters. Especially the run number was used for storing run relevant DCS data. In combination with a shown dynamical integration of more transfer data this was done very efficiently at the combined test beam.\par
For the message transfer we built up a DCS finite state machine to monitor the parameters of the detector modules and to generate corresponding states. Pixel DCS sent messages with severity flags which were read by the shifter during data taking.\par
\begin{figure}[h]
	\centering
		\includegraphics[width=\columnwidth]{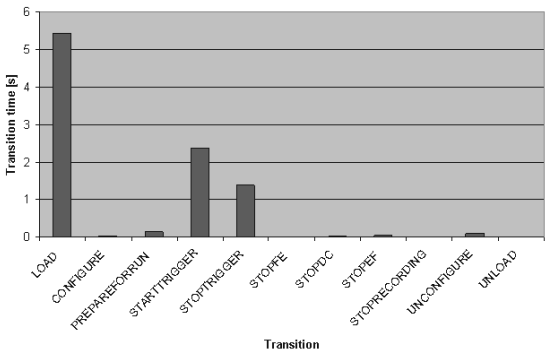}
	\caption{Required time for various transitions}
	\label{fig:Diagramm}
\end{figure}

%%%
Performing timing studies, certain DCS actions were connected to TDAQ transitions. For the reason mentioned above, the interconnection between on- and off-detector parts of the optical link is of special interest. Thus the setting of the reset signal of the SC-OLink which allows a slow and controlled recovery of the delay control circuit was linked to the transition 'LOAD' (see figure \ref{fig:Diagramm}).\par
%%%
%%%
The total time for a transition is composed of the time of the DDC process and the time of the DCS process. For the above example we measured 50 ms for the DDC and around 5 s for the DCS process. Due to these measurements we were able to optimise the control code during the test beam period together with changes in the hardware properties.\par
\begin{figure*}[h]
	\centering
		\includegraphics[width=0.9\textwidth]{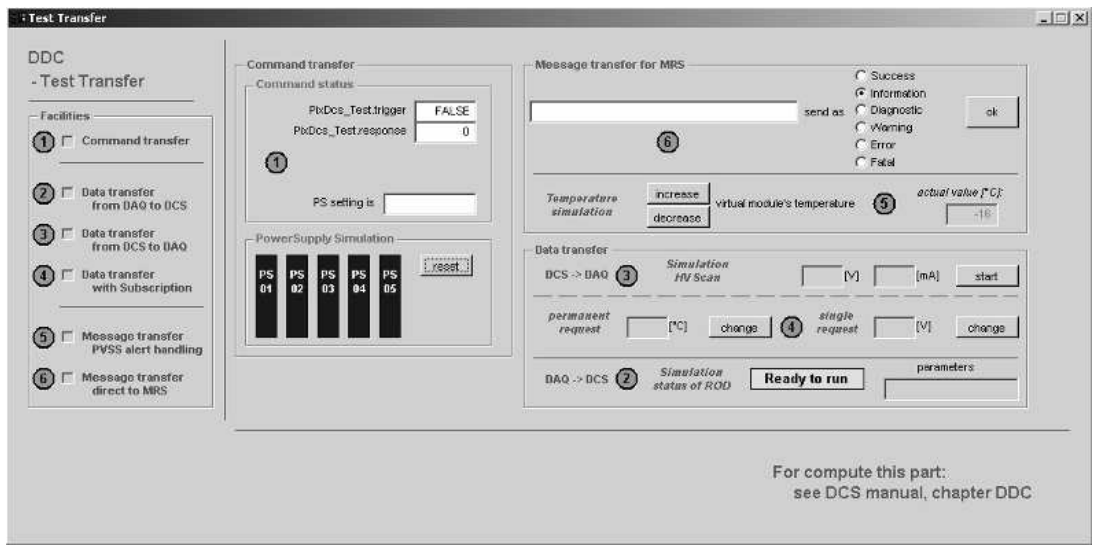}
	\caption{Graphical user interface - DDC}
	\label{fig:ddctest}
\end{figure*}

 
%%%
To verify the full functionality of DDC by the shifter during the experiment, additional tools for data analysing (see figure \ref{fig:ddctest}, last page) are inserted in the structure of the pixel detector control system which did not effect the normal operation. Checking the command transfer is done by switching and setting any number of virtual power supplies while checking the message transfer is done by simulating differently weighted temperatures of a detector module and sending corresponding messages with severity flags. Reviewing the data transfer, one could observe from the TDAQ side a simulated high voltage scan of a virtual detector module inside DCS. On the other hand simulated TDAQ data is visible in DCS.\\
These tools were very helpful during the operation and they are now an inherent part of the detector control system.
%\begin{figure*}[h]
%	\centering
%%		\includegraphics[width=\columnwidth]{corel_bilder/DDCtest.eps}
%		\includegraphics[scale=0.6]{corel_bilder/DDCtest.eps}
%	\caption{Graphical user interface - DDC}
%	\label{fig:ddctest}
%\end{figure*}
%%%  
%%%%%%%%%%%
\section{Interface to the Conditions Database}
\label{section:CondDB}
%%%%%
Conditions data is seen as every data needed for reconstruction besides the event data itself, thus it has to reflect the conditions the experiment was performed and the actual physics data were taken \cite{Amorim1}. For the pixel DCS this includes basically parameters of the detector modules such as voltages, currents and temperatures but also parameters and status information of further DCS devices.\\
As already mentioned, the ATLAS detector control system is based on the software PVSS. PVSS contains an internal archive to store and to read back the history of DCS values, but does not allow to access the data from outside the PVSS framework.\\
Therefore a PVSS API\footnote{\textbf{A}pplication \textbf{P}rogramming \textbf{I}nterface} manager was developed by the Lisbon TDAQ group. This custom made manager is based on a C++ interface between PVSS and a MySQL database. When running, it connects to each of the DCS parameters defined by the user and stores the values together with a timestamp in the database. When a value change occurs, the previous value is updated by replacing the timestamp by a time interval and the new value is stored in the database in the same way as the first value.\par
% \cite{Amorim2}.\\
%%%
\begin{figure}[h]
	\centering
		\includegraphics[width=\columnwidth]{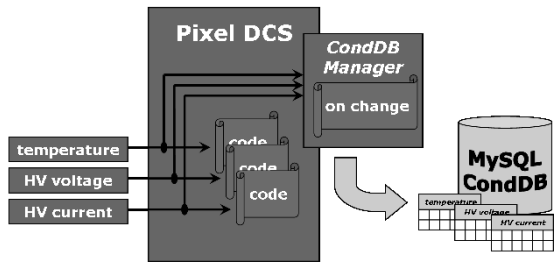}
	\caption{Schematic of data flow through the interface}
	\label{fig:CondDB}
\end{figure}
%%%
During the combined test beam the system reliability of the pixel DCS set up, the amount of data and the handling of the interface have been examinated.\\
Due to the given storing mechanism described above, no data filter or smoothing processes could be used. As one result we had about 5 storing processes per second and per value which produced a non acceptable amount of data and necessitated a limitation of changes. By the integration of a new storing mechanism with a storage at the begin of a run, every minute\footnote{If a monitored value run out of its limits this storage interval was scaled down to get more information about the bahavior} and at the end of the run we were able to reduce the amount of data significantly.\\
Based on about 150 Bytes per storage for a detector module, pixel DCS would produce more than 37 GBytes of data for physics analysis per year at an estimated 5 minutes storage interval.
%%%%%%%%%%
\section{Summary}
\label{summary}
%%%%%
At the combined test beam we have built up a pixel detector control system which worked very well during the four month beam period. Pixel specific software tools were used with good acceptance by shifters. Many functionality issues could be studied sufficiently.\\
The DAQ-DCS communication software was tested intensely and was established very successfully in the pixel configuration. We were able to use the full funcitonality of DDC. We provided commands for several actions inside DCS. TDAQ data were computed by DCS in a well defined way while DCS data was used by TDAQ for monitoring. Messages with severity flags were available. From this point all further requirements to pixel DCS coming with a system scale up could be achieved by this package.\\
DDC is the appropriate tool to handle the interaction between on and off-detector parts of our optical link. It allows us to develop tuning algorithms to find the optimal operation point for the components of the read out chain. As a first step, a graphical user interface which shows inside DCS various parameters of the BOC is currently under development.\\
The used interface to the conditions database did not cover all the pixel DCS aims. After the combined test beam ATLAS intended to use the LHC Computing Grid (LCG) framework for developing a new interface to the conditions database which makes available general database tools and interfaces for subsequent analysis. Further better cofigurability and more flexibility for filtering data as well as the possibility to read data from the conditions database in PVSS has to be considered.
%%%%
\section{Acknowledgments}
These studies are a result of sharing knowledge in the ATLAS pixel DAQ group and the ATLAS DCS group. We would like to thank all people being involved in the  work, especially V. Khomoutnikov for support during the test beam period. He was always open for discussions and gave us a lot of fruitful hints.
% \appendix
% \section{}
% \label{}

%%%
%%%
%Assigning a temperature channel (NTC type), SIT shows the channels available, while grayed out channels are already virtually connected.\\
%%%
\end{document}